\documentclass[prx,aps,letterpaper,twocolumn,allowfontchageintitle=true,unpublished]{quantumarticle}

\usepackage{bbm}
\usepackage{dcolumn}% Align table columns on decimal point
\usepackage{bm}% bold math
\usepackage{graphicx}
\usepackage{hyperref}
\usepackage{color}
\usepackage{version}
\usepackage{mathtools}
\graphicspath{{figures/}}
\usepackage{natbib}
\usepackage{multicol}

\usepackage{float}
\usepackage{wasysym}

\usepackage[table]{xcolor}
\usepackage{booktabs}

\usepackage{xcolor}
\usepackage{braket}

\begin{document}

\title{\centering \color{quantumviolet} How to compute a 256-bit elliptic curve private key \newpage with only 50 million Toffoli gates  \newpage}
\author{\vspace{-5ex}}
\affiliation{Daniel Litinski @ PsiQuantum, Palo Alto}
\date{\vspace{-6ex}}
{\centering \maketitle}

\begin{abstract}

We use Shor's algorithm for the computation of elliptic curve private keys as a case study for resource estimates in the silicon-photonics-inspired active-volume architecture. Here, a fault-tolerant surface-code quantum computer consists of modules with a logarithmic number of non-local inter-module connections, modifying the algorithmic cost function compared to 2D-local architectures. We find that the non-local connections reduce the cost per key by a factor of 300-700 depending on the operating regime. At 10\% threshold, assuming a 10-$\mu$s code cycle and non-local connections, one key can be generated every 10 minutes using 6000 modules with 1152 physical qubits each. By contrast, a device with strict 2D-local connectivity requires more qubits and produces one key every 38 hours. We also find simple architecture-independent algorithmic modifications that reduce the Toffoli count per key by up to a factor of 5. These modifications involve reusing the stored state for multiple keys and spreading the cost of the modular division operation over multiple parallel instances of the algorithm.

\end{abstract}

\begin{figure*}[t]
\centering
\includegraphics[width=\linewidth]{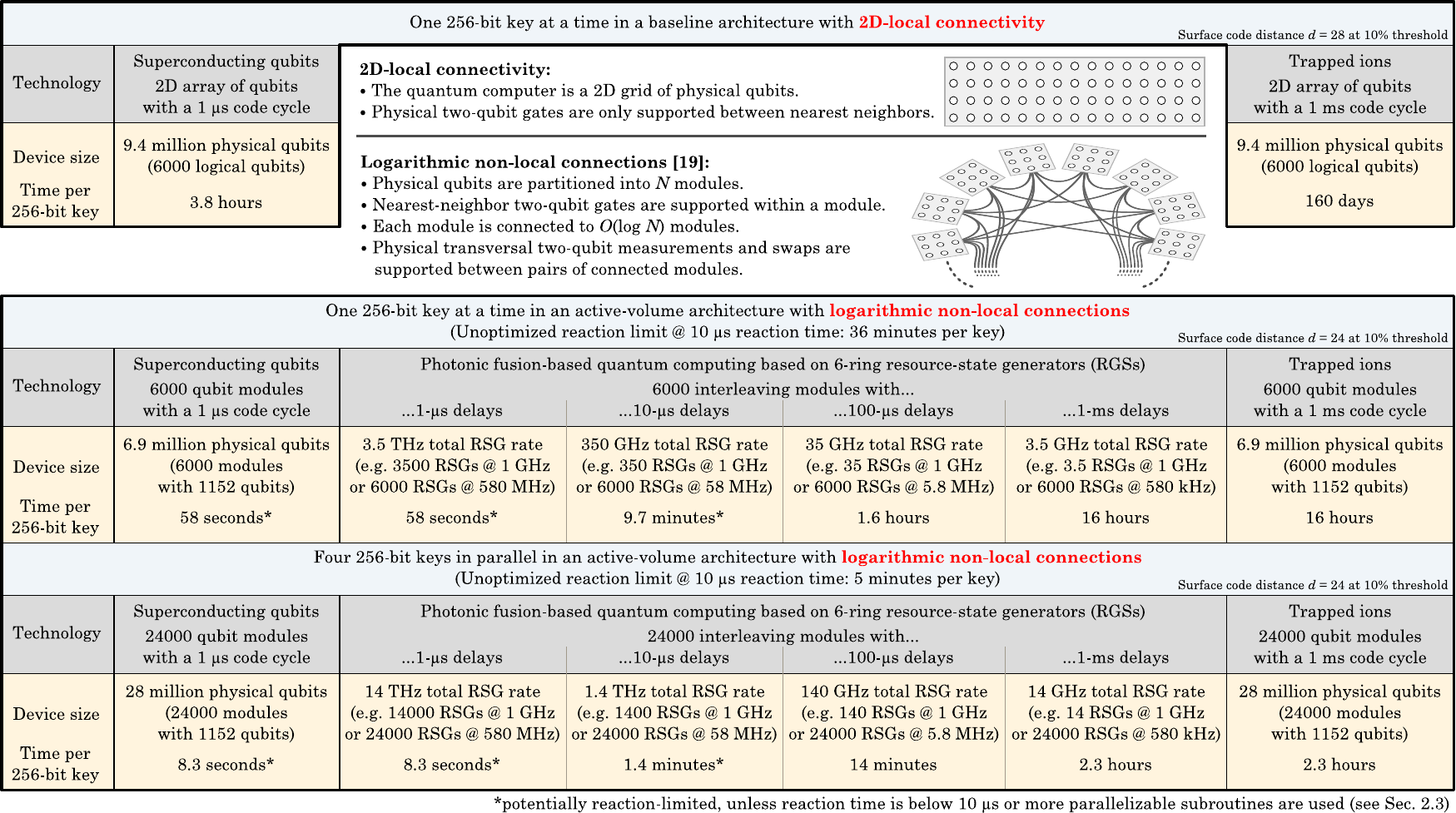}
\caption{Summary of resource estimates performed in Sec.~\ref{sec:resourceestimate}. Note that, for photonic fusion-based quantum computing, longer delays are strictly beneficial in an active-volume architecture with logarithmic non-local connections. Doubling the delay length doubles the memory provided by each RSG, and also doubles the code cycle, but compensates this slowdown by doubling the number of logical operations that are executed in parallel. The additional memory can be used to reduce the cost per key through the cheaper modular inverse operation described in Sec.~\ref{sec:subroutinebreakdown}, and the longer code cycles help avoid the reaction limit.}
\label{fig:summary}
\end{figure*}

The most widely adopted asymmetric cryptography schemes are RSA and elliptic curve cryptography (ECC). The NIST-recommended minimum key size for RSA is 2048 bits, whereas only 256 bits are recommended for ECC~\cite{Barker2009,Chen2023}. This choice is motivated by the resilience of ECC keys against classical-computing-based attacks. However, both cryptosystems are susceptible to the potential threat of quantum computing~\cite{Shor1994,Shor1997}. In this regard, the smaller key size of ECC keys renders them more vulnerable to quantum computers, as substantiated by lower gate counts observed in the existing literature~\cite{Proos2003, Roetteler2017, Haener2020, Gouzien2023}. Consequently, it is reasonable to anticipate that 256-bit ECC will be the first widely used cryptosystem compromised by quantum computing.

What size does a quantum computer need to be to break 256-bit ECC keys, and how much time does it take per key? The problem size surpasses the capability of quantum computers without error correction, necessitating fault-tolerant quantum computing (FTQC)~\cite{Campbell2016,TerhalRMP}. We focus on resource estimates for surface-code-based FTQC~\cite{Kitaev2003,Bravyi1998,Fowler2012}, where logical qubits are encoded as surface-code patches consisting of hundreds or thousands of physical qubits. There exist two types of general-purpose architectures for surface codes: baseline architectures with nearest-neighbor logical two-qubit operations on a 2D grid~\cite{Litinski2019,Fowler2018,Chamberland2022,Chamberland2022a,Bombin2021}, and the recently introduced active-volume architecture~\cite{Litinski2022} utilizing a logarithmic number of non-local connections between patches. The latter leverages non-local connections to parallelize the execution of logical operations, resulting in a significant speedup compared to a fault-tolerant quantum computer with the same footprint, but strict 2D-local connectivity.

\textbf{Different architectures.} 
The existing literature on FTQC resource estimates primarily focuses on baseline architectures, relying on determining logical qubit counts ($n_Q$) and Toffoli gate counts ($n_{\rm Tof}$) due to the relevance of the cost function $n_Q \cdot n_{\rm Tof}$. Resource estimates for active-volume architectures are different, albeit not more complicated. Instead of counting qubits and gates, different fundamental subroutines are counted based on their specific costs, known as the active volume. This paper aims to guide the reader through a simplified active-volume resource estimation procedure using the ECC algorithm as a case study. In the process, we improve architecture-independent gate counts and discuss tailored optimization techniques for active-volume architectures.

The time scale of surface-code quantum computers is defined by the code cycle length $t_C$. The code distance $d$ determines the logical qubit size as $d^2$ physical data qubits and logical cycle duration as $t_L = d \cdot t_C$. In baseline architectures, a quantum computer with the capacity to execute $n_Q$-qubit computations consists of $2n_Q$ logical qubits. Gates are executed sequentially, with one $T$ gate per logical cycle or one Toffoli gate per four logical cycles. In an active-volume architecture, a quantum computer consists of modules with $d^2$ physical data qubits. Each module either operates as a memory or workspace module. Memory modules increase the memory capacity by one logical qubit, while workspace modules enhance computational speed by one \textit{block} per logical cycle. Subroutine costs are measured in \textit{blocks}. Non-local connections between modules enable parallelized logical operations. The number of blocks executed per logical cycle equals the number of workspace qubits. The assignment of memory and workspace qubits can be changed dynamically during runtime. For simplicity, our resource estimates assume an equal allocation of memory and workspace qubits, with $n_L$ logical qubits resulting in $n_L/2$ memory qubits and a speed of $n_L/2$ blocks per logical cycle. Active-volume resource estimates focus on counting the total number of blocks in a computation, i.e., the active volume.

\begin{figure*}[t]
\centering
\includegraphics[width=0.9\linewidth]{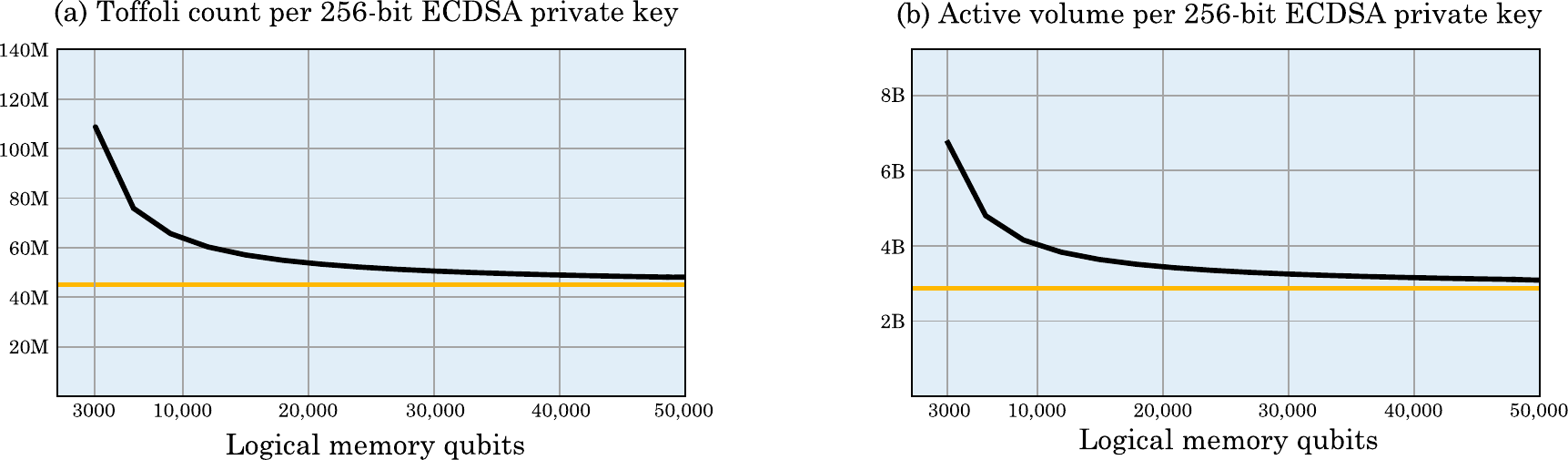}
\caption{Toffoli counts and active volume estimates obtained in this paper. Every instance of the algorithm uses 3000 logical memory qubits. With $k$ instances running in parallel, the Toffoli count per key is $(44 + 65/k)$ million, and the active volume per key is $(2.8 + 3.8/k)$ billion blocks. In the baseline and active-volume architectures considered in this paper, the total number of logical qubits is twice the number of logical memory qubits. The orange lines show the asymptotic Toffoli count and active volume.}
\label{fig:overview}
\end{figure*}

\textbf{Different hardware.} We perform resource estimates for different hardware platforms, namely superconducting qubits, trapped ions, and photonic fusion-based quantum computers (FBQC)~\cite{Bartolucci2021}. For superconducting qubits and trapped ions, we assume circuit-based quantum computers comprising collections of physical qubits executing single-qubit and two-qubit gates. We consider different code cycle lengths (1~$\mu$s for superconducting qubits and 1~ms for trapped ions) to account for the distinct physical time scales. 

In contrast, FBQC is not constructed as an array of physical qubits but rather as a network of chips that generate resource states~\cite{Bartolucci2021,Bombin2021}. How does one then quantify the size of such a device? Resource-state generators (RSGs), responsible for generating specific multi-photon states, form the majority of the device footprint. Additional components for photon rerouting, measurement, and delay lines also contribute to the computation. Therefore, the physical size of the FBQC depends on the total RSG rate ($f_{\rm RSG}$). Note that this rate is unrelated to the physical clock rate, as, e.g., either 350 RSGs clocked at 1 GHz or 6000 RSGs clocked at 58 MHz result in the same total RSG rate of 350 GHz.

While the ratio of logical qubits to physical qubits in circuit-based quantum computers is determined by the code distance, photonic FBQC allows for further flexibility captured by an additional parameter: the temporal length of the longest delay line ($t_d$). Delay lines are devices that store photons for a fixed amount of time. Photons produced by RSGs are only present for a small fraction of the total computation, with the longest-lived photons in the device surviving for $t_d$ between the time they are generated and the time they are measured in a single-photon detector. The delay line length $t_d$ does not limit the maximum duration of the computation, but sets the code cycle time as $t_C = t_d$. 

The number of logical qubits ($n_L$) in an FBQC depends on the maximum number of resource states present simultaneously, which is the total RSG rate multiplied by $t_d$. As each logical qubit has a footprint of $d^2$ resource states in an example device based on 6-ring resource states~\cite{Bartolucci2021, Bombin2021}, $n_L = f_{\rm RSG} \cdot t_d / d^2$.
Longer delays increase the code cycle length and the number of logical qubits contributed by each RSG. The maximum usable delay line length is limited by the transmission loss rate. We consider delay line lengths from 1 $\mu$s to 1~ms to capture the interval between fast superconducting qubits and slow ion traps. Examples of physical instantiations of such delays are fiber delays that are 200 m (1 $\mu$s) or 2 km (10~$\mu$s) long, or free-space delays with mirrors separated \linebreak by 300 meters and 100 reflections (100 $\mu$s) or 1000 reflections (1~ms). Other implementations are also possible.

\subsection*{Overview of results}

\textbf{Algorithmic modifications.}
We begin by breaking down the ECC algorithm into fundamental arithmetic subroutines in Sec.~\ref{sec:subroutinebreakdown}. We introduce three modifications: (1) reusing the state computed in the first half of the algorithm to generate multiple keys by repeating the second half, reducing the cost by up to a factor of 2, \linebreak (2) determining 48 of the 256 bits of the key through brute force search on a classical computer, slightly decreasing the cost, and (3) adapting a classical method~\cite{brent2010modinverse} for computing multiple modular multiplicative inverses using a single inversion and a few multiplications.

The dominant cost in the ECC algorithm is the computation of inverses. It is possible to compute the inverses of multiple numbers $x_1, \dots, x_n$ by first computing their product, then computing the inverse $(x_1 \cdots x_n)^{-1}$ and finally obtaining the individual inverses ${x_i}^{-1}$ through multiplication. We can make use of this by running multiple instances of the ECC algorithm in parallel computing different private keys. When the different instances reach the point when they need to compute an inverse, this method enables them to share resources and use a single inversion to compute the inverses with a reduced cost per instance. Asymptotically, this replaces the cost of modular inversion by the cost of three modular multiplications, which reduces the cost per key by up to a factor of around 2.5.

The resulting behavior is illustrated in Fig.~\ref{fig:overview}, where the Toffoli count and active volume per key decrease with increased quantum computer memory. In a baseline architecture, this is not a favorable trade-off, as doubling the qubit count leads to a less than twofold decrease in the Toffoli count, resulting in an overall increase in the cost function $n_Q \cdot n_{\rm Tof}$. In contrast, active-volume architectures benefit from the reduced Toffoli count and active volume, making them more efficient in a better-than-linear manner with increased size.

\begin{figure}[t]
\centering
\includegraphics[width=\linewidth]{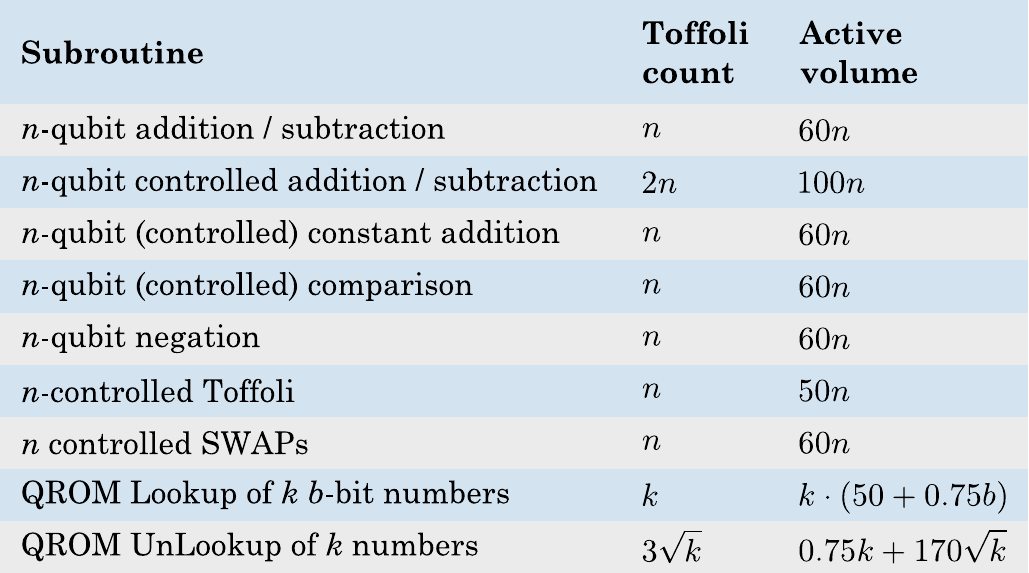}
\caption{Approximated Toffoli counts and active volumes of fundamental subroutines used in this paper. These values are based on the estimates in Ref.~\cite{Litinski2022}.}
\label{fig:basicsubroutines}
\end{figure}

\textbf{Resource estimates.}
Our findings are summarized in Fig.~\ref{fig:summary}. The resource estimate is based on approximated Toffoli counts and active volumes of fundamental arithmetic subroutines and lookup tables, as shown in Fig.~\ref{fig:basicsubroutines}. These estimates are derived from the values presented in Ref.~\cite{Litinski2022}, with a slight overestimation of Toffoli counts and active volumes for simplicity. Additionally, we assume an overestimated qubit count by considering 3000 logical memory qubits per algorithm instance. The numbers presented in Fig.~\ref{fig:summary} assume a physical error rate at 10\% of the surface-code threshold error rate, which is a common assumption in the literature. A more conservative assumption closer to 50\% of the threshold could lead to a doubling of the code distance, resulting in a twofold increase in the computational time and a fourfold increase in the device footprint. Note that there are more optimistic estimates in the literature, considering shorter code cycles and lower-distance surface codes~\cite{Gouzien2023}. However, in this study, we aimed for a more conservative approach.

While a device with a baseline architecture with 6000 logical qubits and a 1 ms code cycle takes 160 days to generate a 256-bit key, a device with an active-volume architecture can generate one key every 16 hours using 25\% less footprint. Devices with shorter code cycles generate keys at a correspondingly faster rate and attain the same 240-fold speedup due to non-local connections.
The source of the footprint reduction is a distance reduction from $d=28$ to $d=24$ due to the reduced volume of the computation. The speedup is primarily derived from the parallel execution of Toffoli gates. While the baseline architecture executes one Toffoli gate every four logical cycles, the 3000 workspace qubits in the active volume architecture execute around 50 Toffoli gates in every logical cycle, if these Toffoli gates are part of arithmetic circuits.

In addition, the active-volume architecture can benefit from the cheaper inversion operation. With a 10-$\mu$s code cycle, a device with 6000 logical qubit modules can generate one key every 9.7 minutes. A device with four times the footprint that generates four keys in parallel can generate one key every 1.4 minutes, i.e., almost 7 times faster, resulting in an overall spacetime volume reduction. More precisely, the device generates four keys simultaneously in 5.6-minute bursts. Photonic implementations of active-volume architectures benefit greatly from longer delay lines, as this increases the number of logical qubits that each RSG provides, which in turn unlocks active-volume reductions reducing the cost per key. With long delay lines, even a device with a relatively small footprint can compute keys at an acceptable rate.

The minimum duration of a fault-tolerant quantum computation is set by the reaction depth multiplied by the reaction time, a physical time scale related to classical processing and communication. A standard assumption in the literature is 10 $\mu$s, as indicated in Fig.~\ref{fig:summary}, but this is not a physical limit, and improvements are possible. We have not focused on optimizing the reaction depth in this study, but further details are provided in Sec.~\ref{sec:reactiondepth}. The reaction limit is particularly important for superconducting qubits with short code cycles or high-footprint photonic devices with short delay lines.

\begin{figure*}[t]
\centering
\includegraphics[width=0.9\linewidth]{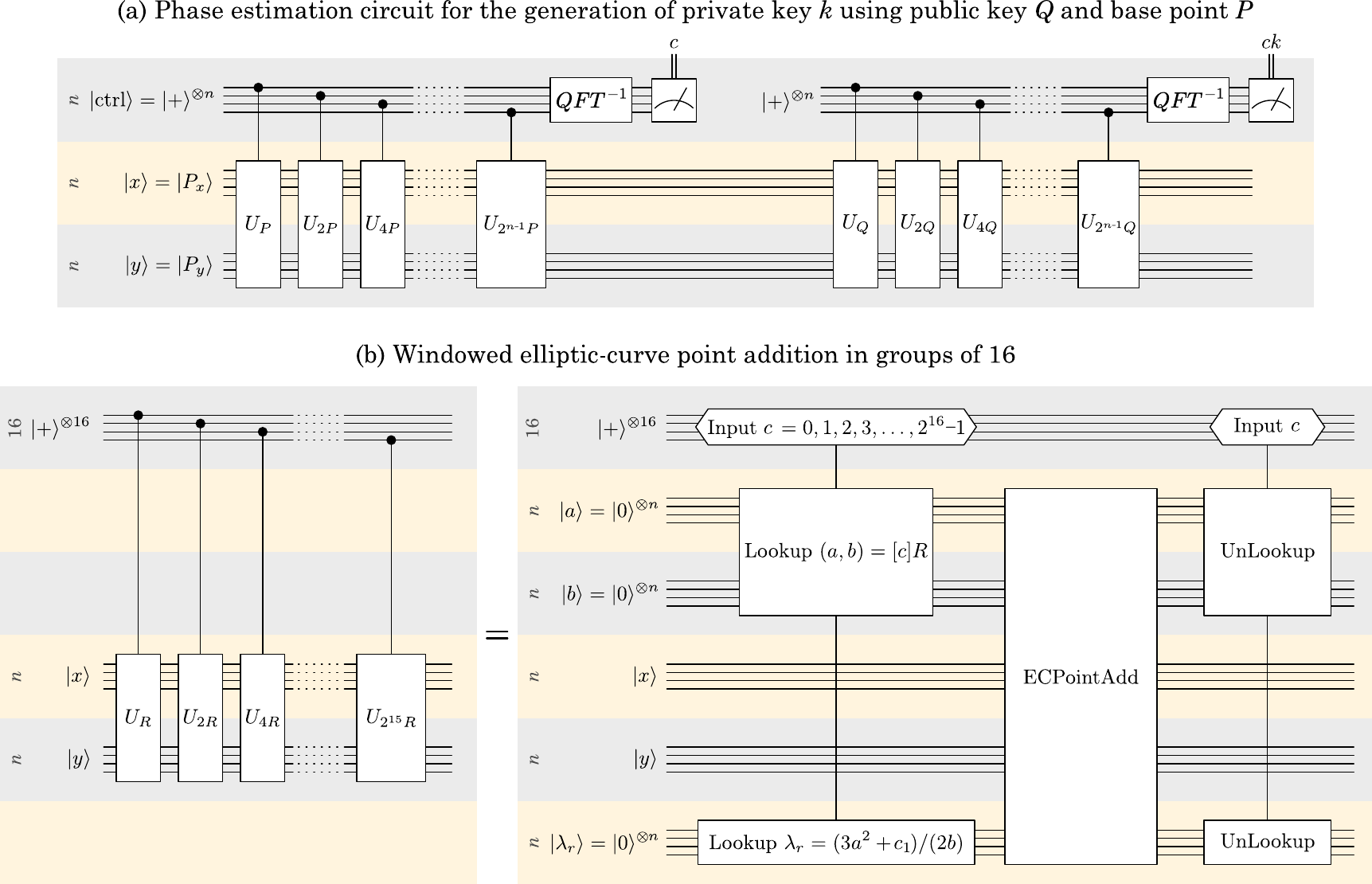}
\caption{(a) Basic structure of the algorithm~\cite{Shor1994, Shor1997} composed of two phase estimations. (b) Groups of 16 controlled unitaries are implemented using a lookup table, an elliptic curve point addition and an uncomputation of the lookup table, following the windowing technique described in Refs.~\cite{Haener2020}. A windows size of 16 was chosen to optimize the resource count for 256-bit keys. The optimal window size may differ for larger keys.}
\label{fig:qpecircuit}
\end{figure*}

\section{Subroutine breakdown}
\label{sec:subroutinebreakdown}

In this section, we decompose the ECC algorithm into the fundamental subroutines shown in Fig.~\ref{fig:basicsubroutines} and perform a Toffoli count and active volume estimate based on the estimates provided in Ref.~\cite{Litinski2022}. Note that Ref.~\cite{Litinski2022} only provides estimates for $n$-qubit adders, Toffolis, SWAPs and QROM lookups. The remaining entries in Fig.~\ref{fig:basicsubroutines} are derived from these values, as subtraction can be performed by flipping all bits of the subtrahend and performing an addition with the input carry set to 1, comparison can be performed using a subtraction, and negation can be performed by flipping all bits followed by an increment operation. The UnLookup refers to the uncomputation of a lookup table using controlled SWAPs and a smaller lookup table as described in Ref.~\cite{Gidney2019b}. We will apply the cost estimate for the $n$-controlled Toffolis to Toffoli gates with $n$ controls as well as groups of $n$ standard 2-controlled Toffoli gates.

\textbf{Elliptic curve keys.} The overall goal of the quantum algorithm is to determine a private key $k$ using the public key $Q$ as an input. In elliptic curve cryptography, a cryptographic scheme is defined via an elliptic curve
\begin{equation}
	y^2 = x^3 + c_1 x + c_2
\end{equation}
with curve parameters $c_1$ and $c_2$, as well as a (typically prime) modulus $p$ and a base point $P = (P_x, P_y)$. Points on this curve are pairs of integer coordinates modulo~$p$. A key pair can be created by generating a random integer $0 \leq k \leq p-1$ as the private key and computing $Q = [k]P$ as the public key via elliptic curve point multiplication. Given two elliptic curve points $P_1 = (a,b)$ and $P_2 = (x,y)$, the addition operation $P_3 = P_1 + P_2 = (x_r, y_r)$ is defined as:
\begin{multline}
	(x_r,y_r)=
	\begin{cases}
		\O & \text{if $P_1 = -P_2$},\\
		(x_r,y_r) = (x,y) & \text{if $P_1 = \O$},\\
		(x_r,y_r) = (a,b) & \text{if $P_2 = \O$}
	\end{cases} \\
\text{else} \\
x_r = \lambda^2 - x - a \mod p \\
y_r = \lambda (a-x_r) - b \mod p \\
\text{with} \quad \lambda = 
\begin{cases}
\lambda_r = \frac{3a^2+c_1}{2b} = \frac{y_r+b}{a-x_r} & \text{if $P_1 = P_2$} \\
\frac{y-b}{x-a} = \frac{y_r+b}{a-x_r} & \text{else}
\end{cases}
\label{eqn:addition}
\end{multline}
Here, $-P_2 = (x,-y)$, and $\O$ refers to the point at infinity, for which we will choose the representation $\O \equiv (0,0)$. Note that all arithmetic operations, including addition, multiplication and division, are modular arithmetic operations modulo $p$. Furthermore, an additional known parameter is the order of a curve $r$, which is defined as $[r]P = P$. A multiple of the base point $P$ can be computed efficiently via repeated doubling and addition. However, there is no known efficient classical algorithm for the reverse operation. Given a public key as a multiple of the base point $Q = [k]P = P + P + \cdots + P$, classical computers are currently unable to compute $k$ efficiently, which is the primary feature exploited in elliptic curve cryptography.

\begin{figure}[t]
\centering
\includegraphics[width=\linewidth]{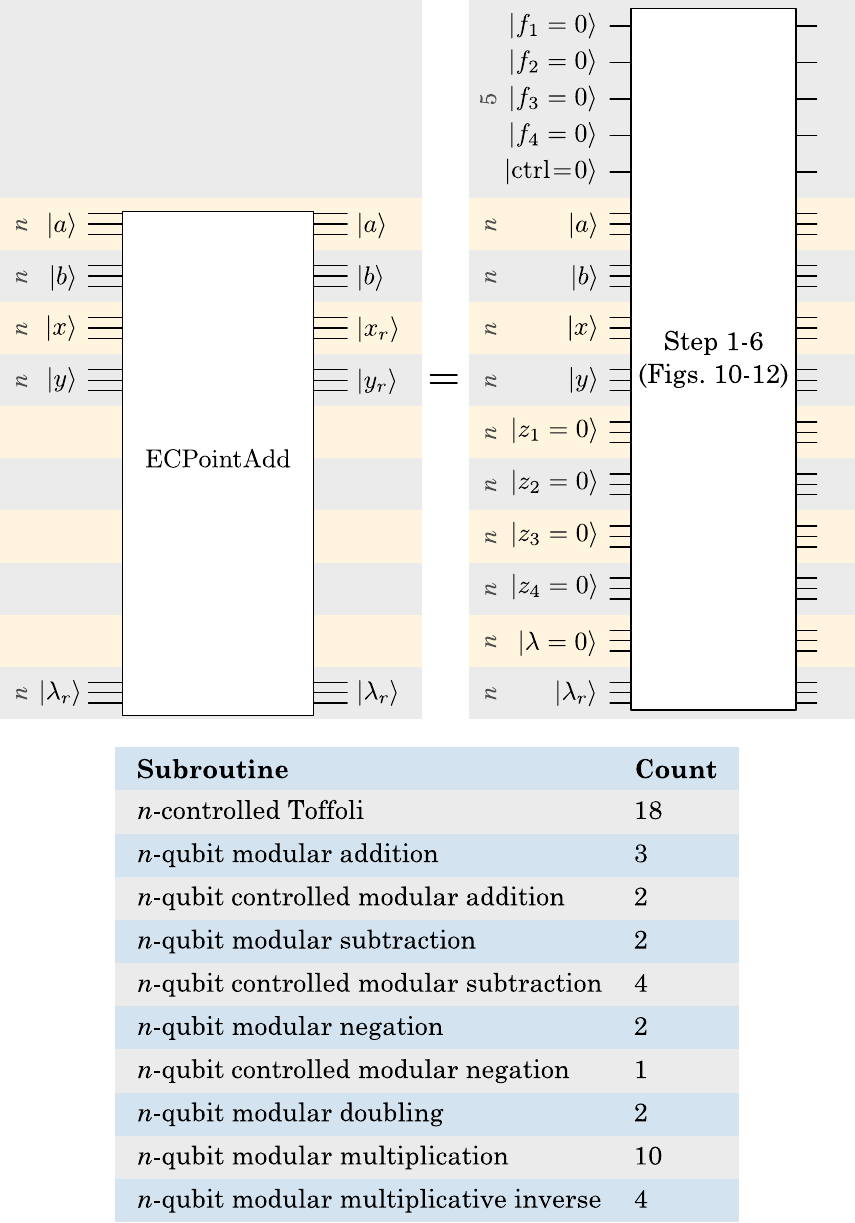}
\caption{Subroutine breakdown of the elliptic curve point addition operation described in Figs.~\ref{fig:ecpointadd1}-\ref{fig:ecpointadd3}.}
\label{fig:ecpointaddfull}
\end{figure}

\begin{figure*}[t]
\centering
\includegraphics[width=\linewidth]{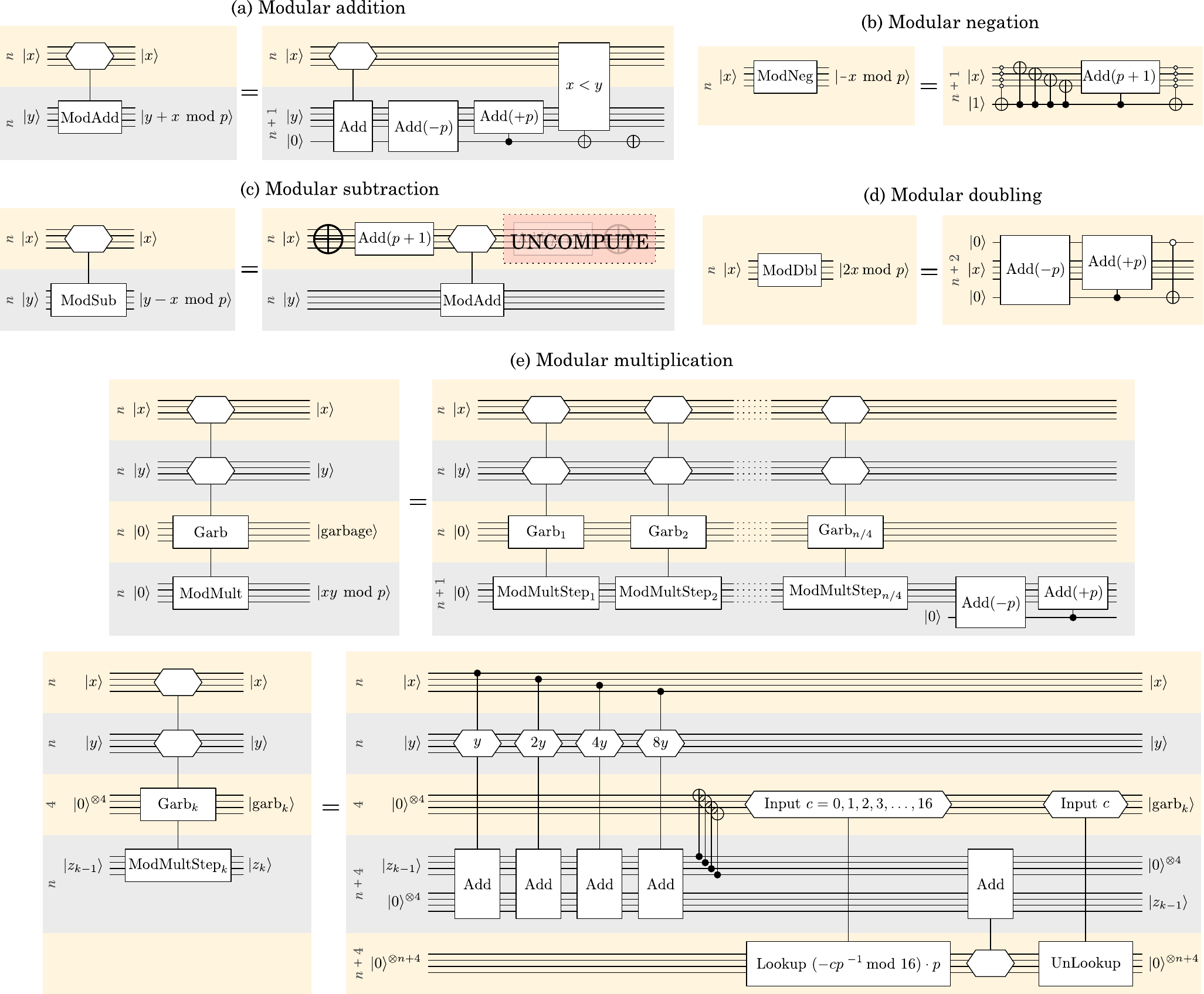}
\caption{Modular arithmetic operations described in Refs.~\cite{Haener2020, Gouzien2023}.}
\label{fig:modulararithmetic1}
\end{figure*}

\begin{figure*}[t]
\centering
\includegraphics[width=\linewidth]{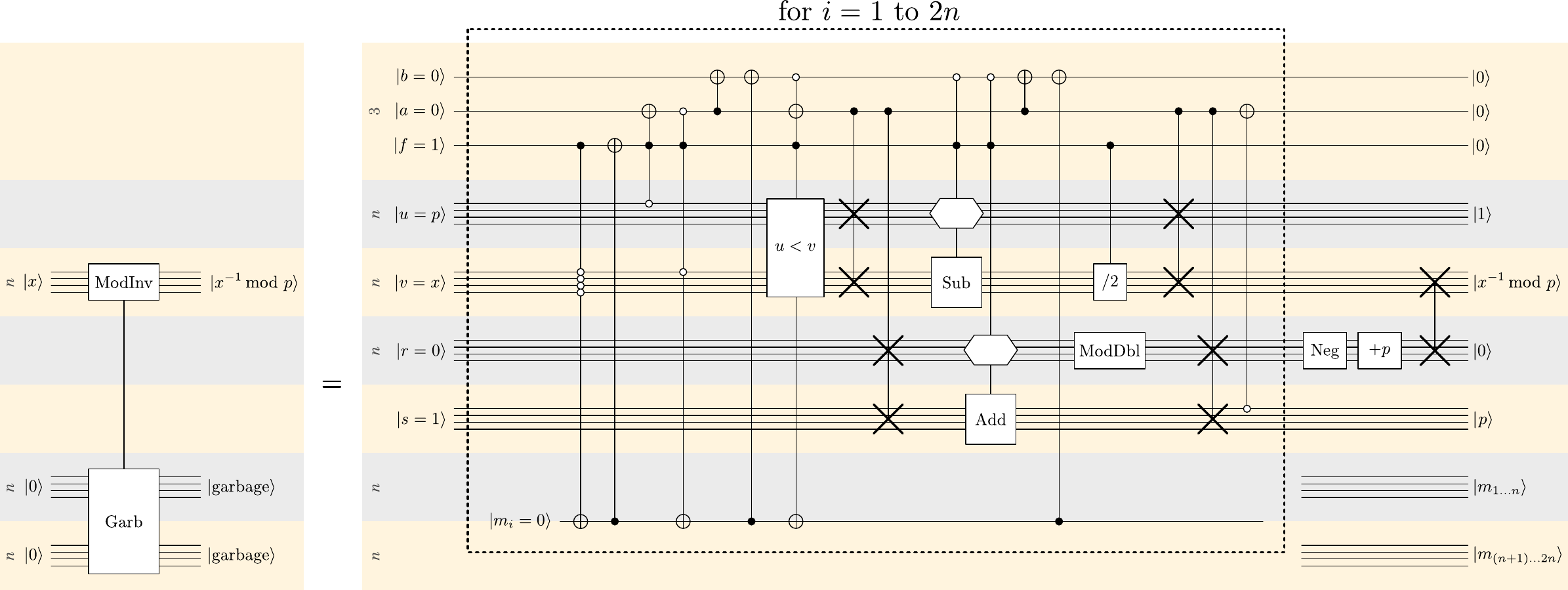}
\caption{Efficient modular inversion circuit described in Ref.~\cite{Gouzien2023}.}
\label{fig:modulararithmetic2}
\end{figure*}

\begin{figure}[b!]
\centering
\includegraphics[width=\linewidth]{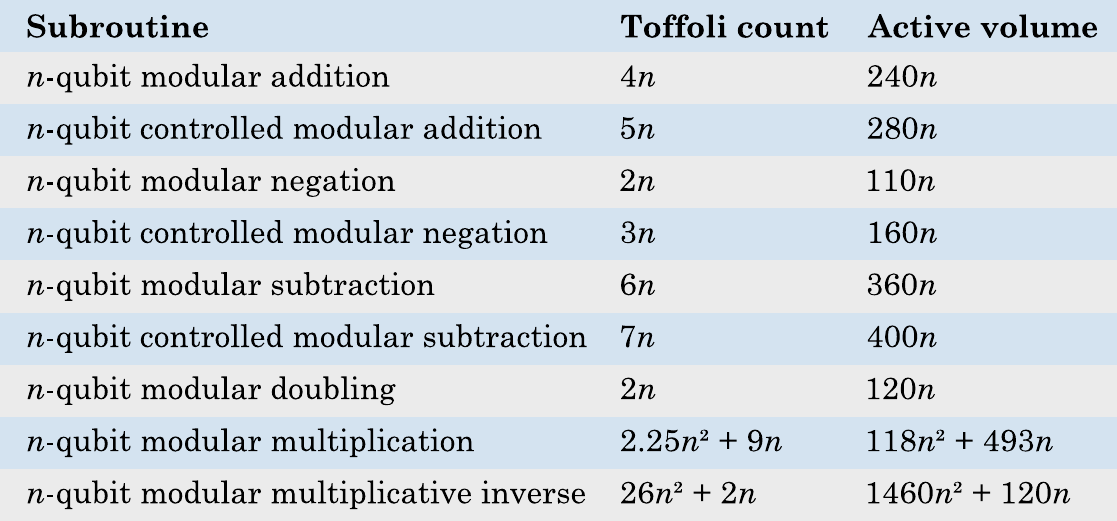}
\caption{Summary of resource estimates for modular arithmetic operations.}
\label{fig:arithmeticresources}
\end{figure}

\textbf{Overall structure.} A quantum computer can compute $k$ efficiently using Shor's algorithm for elliptic curve discrete logarithms~\cite{Shor1994, Shor1997}. The overall structure of this algorithm is shown in Fig.~\ref{fig:qpecircuit}a for the computation of $n$-bit keys. Note that we use gray numbers to the left of qubit registers to indicate the number of qubits in each register. The algorithm consists of two phase estimation steps with unitaries $U_P$ and $U_Q$ acting on two $n$-qubit registers $|x\rangle$ and $|y\rangle$, where
\begin{equation}
	U_P |R = (x,y)\rangle = |R + P\rangle
\end{equation}
and
\begin{equation}
	U_Q |R\rangle = |R + Q\rangle = |R + [k]P\rangle
\end{equation}
are unitary operators performing elliptic curve point addition with the base point $P$ and public key $Q$, respectively. The phase estimation is performed on an input state $|P\rangle = |P_x\rangle \otimes |P_y\rangle$. Note that the eigenstates of $U_P$ that overlap with $|P\rangle$ are all of the form
\begin{equation}
	|\psi_c\rangle = \sum\limits_{j=0}^{r-1} e^{i c j \frac{2\pi}{r}}|[j]P\rangle, \quad U_P|\psi_c\rangle = e^{-ic \frac{2\pi}{r}} |\psi_c\rangle \, .
\end{equation}
Therefore, a quantum phase estimation with $U_P$ in the first half of the algorithm prepares a random state $|\psi_c\rangle$ and generates the corresponding integer $c$ with $0~\leq~c~\leq~r-1$ as an output. The state $|\psi_c\rangle$ is a simultaneous eigenstate of $U_Q$ with
\begin{equation}\begin{split}
 U_Q |\psi_c\rangle &= \sum\limits_{j=0}^{r-1} e^{i c j \frac{2\pi}{r}}|[j]P + Q\rangle \\ &= \sum\limits_{j=0}^{r-1} e^{i c j \frac{2\pi}{r}}|[j+k]P \rangle 
 = e^{-i c k \frac{2\pi}{r}} |\psi_c\rangle \, .
\end{split}\end{equation}
The second phase estimation with $U_Q$ on the state $|\psi_c\rangle$ generates the eigenphase $ck$ as an output. Since $c$ is known from the first phase estimation, $k$ can be obtained after a modular division of $ck$ by $c$.

Naively, each phase estimation requires $n$ repetitions of controlled elliptic curve point additions. The windowing technique introduced in Refs.~\cite{Gidney2019b, Gidney2021, Haener2020} is used to reduce the number of lookup additions, as shown in Fig.~\ref{fig:qpecircuit}b. Each group of 16 controlled unitaries is replaced by a single point addition operation \linebreak (ECPointAdd) and a QROM lookup of $2^{16}$ pre-computed elliptic curve points $(a,b) = [c]R$ for $c \in [0 \dots 2^{16}-1]$ together with the constant $\lambda_r$ required for point doubling, as well as an uncomputation of the lookup table. While the optimal window size will depend on the key size, we found a windows size of 16 to be close to optimal for 256-bit keys. Note that there are 16 such group-of-16 operations in each phase estimation step, where $R$ takes the values $R=2^{16j}P$ for $0 \leq j \leq 15$.

\begin{figure*}[t]
\centering
\includegraphics[width=0.98\linewidth]{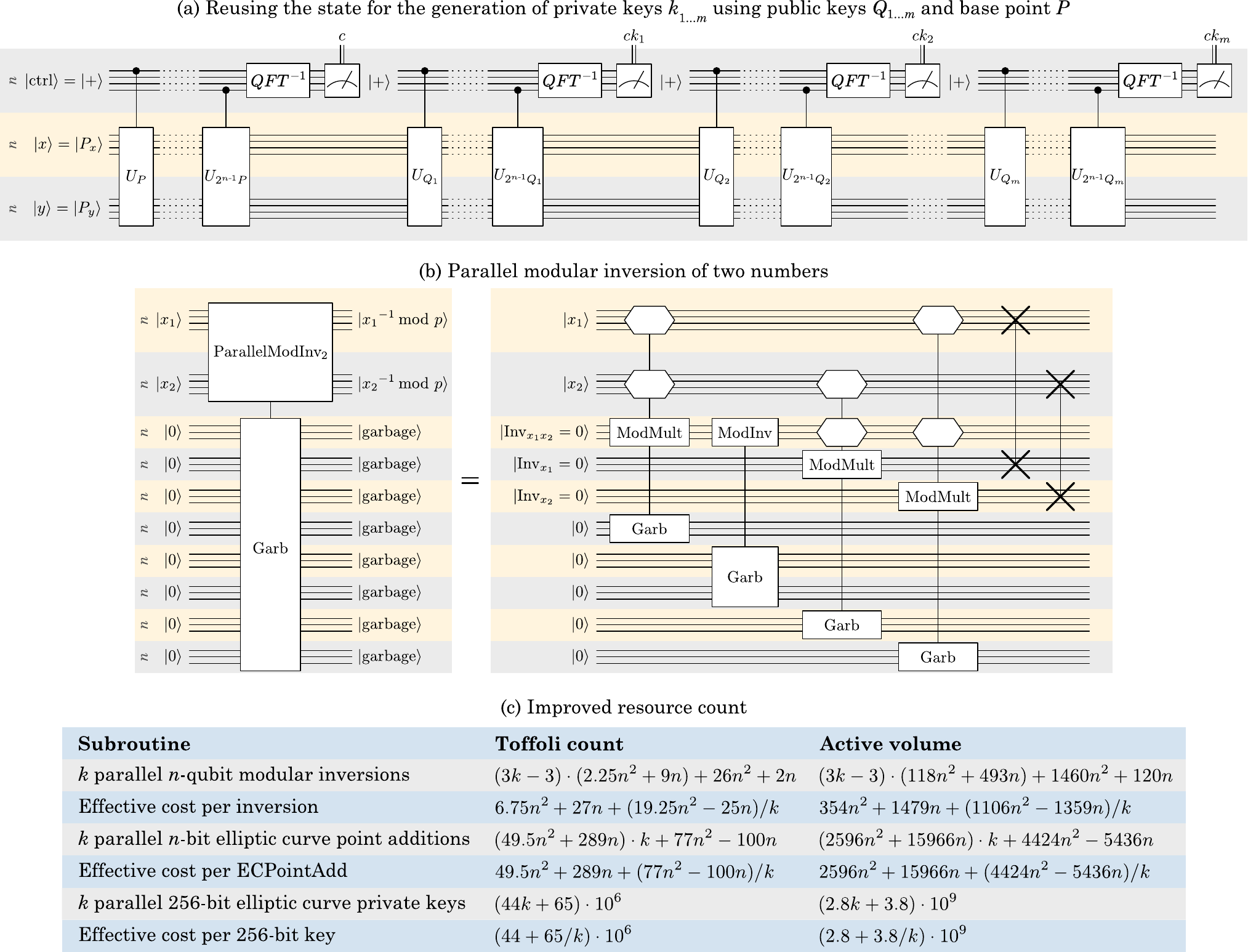}
\caption{Algorithmic modifications. (a) Multiple keys can be generated by repeating the second half of the algorithm for different public keys. (b) $k$ parallel modular inversions can be performed with $3k-3$ modular multiplications and a single modular inversion. (c) Resulting resources count after the algorithmic modifications.}
\label{fig:improvements}
\end{figure*}

\textbf{Elliptic curve point addition.} Our construction of the ECPointAdd circuit is mostly based on the construction presented in Ref.~\cite{Roetteler2017}. The main difference is that we construct an exact circuit for the point addition operation, whereas the circuit in Ref.~\cite{Roetteler2017} ignores the exceptional cases where input or output points are $\O$, or point doubling is performed. This slightly increases the cost of the point addition operation, but since we will later consider performing multiple repetitions of this operation when generating multiple keys, we may be less tolerant to algorithmic errors.

The circuit shown in Fig.~\ref{fig:ecpointaddfull} inputs two points $P_1~=~(a,b)$ and $P_2 = (x,y)$ stored in four $n$-qubit registers, as well as the constant $\lambda_r = (3a^2 + c_1)/(2b)$. It performs an in-place point addition using $5n+5$ additional ancilla qubits and outputs the result in the $(x,y)$ register. The first 5 ancilla qubits $|f_{1-4}\rangle$ and $|\mathrm{ctrl}\rangle$ are used to indicate the exceptional cases of the point addition. The flags $f_1$ - $f_4$ indicate that $a = x$, $b = -y$, $P_1 = \O$ and $P_2 = \O$, respectively.

In step 1 of the circuit shown in Fig.~\ref{fig:ecpointadd1}, we first set these flags by using equality checks between $n$-qubit registers, as well as modular negations and $n$-qubit Toffolis. Each $n$-qubit equality check can be implemented with an $n$-qubit Toffoli conjugated by CNOT gates, so we assign a cost of 6 $n$-controlled Toffoli gates and 2 $n$-qubit modular negations to step 1. The $\mathrm{ctrl}$ flag is set if $f_2 = f_3 = f_4 = 0$. This flag determines that neither the input points nor the output point are $\O$, indicating that we are in the \textit{else} branch of the point addition described in Eq.~\ref{eqn:addition}. Note that we ignore negligibly cheap operations such as the 3-controlled Toffoli used at the end of step 1. In step 2, we compute $\lambda$ into the last ancilla register, either via modular arithmetic if $f_1 = 0$, or by copying $\lambda_r$ into the register if $f_1=1$. An equality check between $\lambda$ and $\lambda_r$ resets the flag $f_1$. Note that a red uncompute box is meant to describe the complex conjugate of the operation contained inside the box.

In steps 3 and 4 (Fig.~\ref{fig:ecpointadd2}), we use modular arithmetic operations to compute $a-x_r$ and $y_r+b$ in the $|x\rangle$ and $|y\rangle$ registers. In step 5 (Fig.~\ref{fig:ecpointadd3}), the $|\lambda\rangle$ register is reset to $|0\rangle$ by exploiting the fact that $\lambda$ can be \linebreak (un-)computed from $a-x_r$ and $y_r+b$.  At the end of step~5, the $|x,y\rangle$ registers either contain $x_r$ and $y_r$ if $\mathrm{ctrl} = 1$, or the unchanged inputs $x$ and $y$ if $\mathrm{ctrl} = 0$, thereby completing the \textit{else} branch of Eq.~\ref{eqn:addition}. In step~6, the $\mathrm{ctrl}$ flag is reset and the exceptional cases are treated. If $f_4=1$, then $P_2 = \O$ and $P_1$ is copied into the output register. The flag is reset via an equality check. If $f_3=1$, then $P_1 = \O$ and no operation needs to be performed, but the flag needs to be reset via a $2n$-qubit Toffoli. Finally, if $f_1 = f_2 = 1$, then $P_1 = -P_2$ and the output needs to be set to $(0,0)$ via a subtraction and an addition before the flag is reset.

The total subroutine count is shown in Fig.~\ref{fig:ecpointaddfull}. In the next step, we decompose these modular arithmetic subroutines into the elementary subroutines listed in Fig.~\ref{fig:basicsubroutines}.

\textbf{Modular arithmetic.}
We decompose modular arithmetic subroutines following the prescriptions in Refs.~\cite{Roetteler2017, Haener2020, Gouzien2023} and perform an active volume estimate. We implicitly assume that all numbers are stored in Montgomery representation~\cite{Montgomery1985} to take advantage of existing efficient constructions for modular multiplication and inversion.

In-place modular addition (Fig.~\ref{fig:modulararithmetic1}a) is implemented by first converting $y$ into an $n+1$-qubit number with the most significant bit initialized as $|0\rangle$ and then performing a non-modular addition. Next, a modular reduction is performed by subtracting the modulus $p$ and checking if the result is negative (if the most significant bit is 1), in which case $p$ is added back to the result. The most significant bit is reset to 0 via a comparator. Using the resource estimates in Fig.~\ref{fig:basicsubroutines} for adders, constant adders and comparators, the Toffoli count of $n$-qubit modular addition is $4n$ and the active volume is $240n$. In a controlled modular addition, only the first addition and the comparator need to be controlled, leading to a Toffoli count of $5n$ and an active volume of $280n$.

Modular negation of $x$ (Fig.~\ref{fig:modulararithmetic1}b) is performed by flipping all bits and adding $p+1$, unless $x=0$, in which case no operation needs to be performed. The exceptional case is checked with a flag that is computed and uncomputed with an $n$-qubit Toffoli. While the circuit shows two $n$-qubit Toffolis, the uncomputation can be done using measurements, if the temporary AND ancilla qubits~\cite{Gidney2018} are kept throughout the operation. The Toffoli count is therefore $2n$ and the active volume $110n$. In a controlled modular negation, the CNOTs flipping the bits are replaced by $n$ Toffoli gates and the addition of $p+1$ is replaced by a controlled addition, increasing the Toffoli count to $3n$ and the active volume to $160n$.

Modular subtraction (Fig.~\ref{fig:modulararithmetic1}c) can be done by performing a modular negation of the subtrahend without checking for the exceptional case of $x=0$, followed by a modular addition and an uncomputation of the negation. The total Toffoli count is  $6n$ and the active volume is $360n$. Controlling the adder to implement a controlled subtraction increases the Toffoli count to $7n$ and the active volume to $400n$.

Modular doubling (Fig.~\ref{fig:modulararithmetic1}d) is performed by converting $x$ to an $n+2$-bit integer by attaching two $|0\rangle$ qubits as the least and most significant bits. A reduction is performed using the same operations as in the modular addition. The least significant bit is reset to 0 by checking only the most significant bit. If a reduction took place (i.e., $2x \geq p$), then the resulting number is odd and the bit does not need to be reset. If no reduction took place (i.e., $2x < p$), then the resulting number is even and the bit is reset with a zero-controlled-NOT gate.

The out-of-place modular multiplication operation shown in Fig.~\ref{fig:modulararithmetic1}e is described in detail in Section 4.1 of Ref.~\cite{Haener2020} and Appendix C4 of Ref.~\cite{Gouzien2023} and relies on the Montgomery representation. It generates an additional garbage register that can be uncomputed whenever the multiplication is uncomputed. The $n$-qubit multiplication consists of $n/4$ steps, each adding 4 qubits to the garbage register. In each step, four controlled additions are performed. An efficient modular reduction together with a division by 16 is performed using a 16-item lookup table and an addition. Therefore, each step has a Toffoli count of $9n + 28$ and an active volume of $472n + 1492$. The $n/4$ steps are followed by a final reduction using two constant adders, resulting in a total Toffoli count of $2.25n^2 + 9n$ and an active volume of $118n^2 + 493n$. Again, the window size of 16 is motivated by 256-bit keys, while the optimal value will depend on the number of bits.

Finally, in Fig.~\ref{fig:modulararithmetic2}, we show the efficient modular inversion circuit described in detail in Appendix C5 of Ref.~\cite{Gouzien2023}, which is an improved version of the circuits in Refs.~\cite{Roetteler2017, Haener2020} based on Kaliski's algorithm~\cite{Kaliski1995}. It computes the modular multiplicative inverse $x^{-1}$ in-place using five $n$-qubit ancilla registers and generating two $n$-qubit ancilla registers as garbage. The four registers labeled $|u\rangle$, $|v\rangle$, $|r\rangle$ and $|s\rangle$ are initialized in $u=p$, $v=x$, $r=0$ and $s=1$. The highlighted block in Fig.~\ref{fig:modulararithmetic2} is repeated $2n$ times. It mainly contains an $n$-controlled Toffoli, a comparator, a controlled addition, a controlled subtraction, a modular doubling, four $n$-qubit controlled SWAPs and a controlled non-modular halving (which can be implemented with $n$ controlled SWAPs), and therefore has a Toffoli count of $13n$ and an active volume of $730n$. Each repetition of this block adds one qubit to the garbage register. After $2n$ repetitions, the terminal values are $u=1$, $v=0$, $r=p-x^{-1}$ and $s=p$. After the final negation, constant addition and SWAP, all registers apart from the output and garbage registers contain known values, and can therefore be discarded. The total cost of this inversion operation is $26n^2+2n$ Toffolis or $1460n^2 + 120n$ blocks of active volume. This operation is over 10 times more expensive than a multiplication, and is therefore the dominant contribution to the cost of elliptic curve point addition. Note that this gate count is slightly lower than the gate count in Ref.~\cite{Gouzien2023} because of the use of Gidney-style arithmetic circuit~\cite{Gidney2018} with fewer Toffoli gates due to measurement-based uncomputation. We summarize all resource estimates of modular arithmetic operations in Fig.~\ref{fig:arithmeticresources}.

\textbf{Algorithmic modifications.} We now introduce three simple modifications to the algorithm to reduce the resource count.

First, we observe in Fig.~\ref{fig:improvements}a that multiple private keys $k_{1\dots m}$ of the same elliptic curve can be generated by repeating the second phase estimation $m$ times for different public keys $Q_{1 \dots m}$, as the state $|\psi_c\rangle$ is a simultaneous eigenstate of all $U_{Q_j}$ of valid public keys $Q_j$. In our resource estimates, we will choose the code distance such that the expected runtime before a logical error is 10 phase estimation blocks. We will terminate the algorithm after a block generates an invalid private key. Therefore, rather than assuming that this method halves the cost per key, we will take into account the initial phase estimation block that generates the state $|\psi_c\rangle$ by increasing the cost per key by a factor of $10/9$ compared to the cost of a single phase estimation block.

For our second modification, we observe that it is not necessary to generate all bits of the private key on the quantum computer. Instead of using 16 repetitions of the group-of-16 operation in Fig.~\ref{fig:qpecircuit}b, we can use only 13 repetitions to generate 208 of the 256 bits of the rescaled private key $ck$. This narrows down the list of possible candidates to $2^{48}$ keys. The correct key can then be found via brute force search. In Ref.~\cite{Courtois2016}, it was shown that a single CPU can test $\mathcal{O}(10^5)$ keys per second. Therefore, a list of $2^{48}$ candidate keys can be searched in $\mathcal{O}(10^6)$ CPU hours. With costs per CPU hour on the order of cents, this can be a worthwhile trade-off reducing the cost per key by a factor of $13/16$.

The third modification aims to reduce the cost of the modular inversion operation. Suppose we are running two instances of the ECC algorithm on the same (larger) quantum computer to break two keys at the same time. We execute the subroutines of the two instances in an alternating fashion~--~first a subroutine of the first instance and then a subroutine of the second instance. Eventually, both instances will reach a modular inversion operation. Rather than performing two costly modular inversion operations, we can adapt a method from classical computation~\cite{brent2010modinverse} to compute both inverses using a single inversion and three multiplications. As shown in Fig.~\ref{fig:improvements}b, we can invert two numbers $x_1$ and $x_2$ by first computing the product $x_1x_2$, inverting the product, and obtaining the individual inverses using two additional multiplications. 

If we have four instances running in parallel, we can invert four numbers using 9 multiplications and one inversion, as shown in Fig.~\ref{fig:parallelmodinv1}. More generally, $k$ numbers can be inverted using $3k-3$ multiplications and a single inversion, or 3 multiplications and $1/k$ inversions per instance. Note that the naive construction uses $6k-4$ $n$-qubit garbage registers, but this can be reduced to $2k + 2\log_2 k$ garbage registers by making inverses available sequentially, as shown in Fig.~\ref{fig:parallelmodinv2}. The asymptotic cost per inversion is three modular multiplications, which reduces the active volume per inversion by up to a factor of around 4, but requires more memory to run multiple instances of the algorithm in parallel.

By adding the costs of all subroutines listed in Fig.~\ref{fig:ecpointaddfull}, we arrive at the cost estimate for an ECPointAdd operation in Fig.~\ref{fig:improvements}c.

\section{Resource estimate}
\label{sec:resourceestimate}
 
The full algorithm to break 256-bit ECC keys consists of 13 repetitions of the group-of-16 operation shown in Fig.~\ref{fig:qpecircuit}b per key. Each operation consists of a $2^{16}$-item table lookup of 768-bit numbers, an ECPointAdd operation, and an uncomputation of the lookup table. The resulting Toffoli count for $k$ parallel instances is $(44k + 65) \cdot 10^6$ Toffoli gates and $(2.8k + 3.8) \cdot 10^9$ blocks of active volume. Asymptotically, the cost per key approaches 44 million Toffoli gates and 2.8 billion logical blocks of active volume for large quantum computers. We adjust these costs by a factor of $10/9$ to account for the expected time before a logical error of 10 phase estimation blocks.  

The circuits shown in Figs.~\ref{fig:qpecircuit} and \ref{fig:ecpointaddfull} contain 11 registers of 256 qubits each. Note that the size of the control register in the phase estimation can be reduced substantially by using iterative phase estimation. For simplicity, we will use an overestimate of 3000 logical memory qubits per instance of the ECC algorithm. This can account for additional ancilla qubits that may be present during the execution of arithmetic circuit in the baseline architecture, or stale magic states and bridge qubits stored in memory in an active-volume architecture~\cite{Litinski2022}.
 
\subsection{Baseline architecture}

We first perform a resource estimate for a baseline architecture with 2D-local connections. This architecture does not benefit from the parallel modular inverse operation, so we only consider one instance of the algorithm. In total, we need 6000 logical qubits and 109 million Toffoli gates per key.

\textbf{Determining the code distance.} With four logical cycles per block, each key generation has a spacetime volume of $2.6 \times 10^{12}$ logical blocks of size $d^3$. Each such block accounts for a $d \times d$ patch of physical data qubits operating for $d$ code cycles and can be estimated to contribute a logical error rate of $p_L = 10^{-d/2}$ at 10\% threshold~\cite{Bombin2021a}. We choose the code distance such that the total error probability after 10 phase estimation blocks (i.e., after executing $2.6 \times 10^{13}$ logical blocks) is slightly below 50\%. With $d=28$, the probability of a logical error after 10 phase estimation blocks is around 20\% in this estimate. Taking into account the factor of $10/9$, the average time per key is 484 million logical cycles, where a logical cycle consists of $d$ code cycles.

\textbf{Circuit-based quantum computers.} For superconducting qubits and trapped ions, each distance-$d$ logical qubit consists of $2d^2$ physical qubits, taking into account the ancilla qubits required for the measurement of surface-code check operators. Therefore, both require 9.4 million physical qubits. With a 1 $\mu$s code cycle, superconducting qubits generate one key every $484 \cdot 28$ seconds, or 3.8 hours. With a 1 ms code cycle, trapped ions generate one key every 160 days.

\textbf{Photonic FBQC.} In order to encode 6000 logical qubits in a photonic fusion-based quantum computer based on 6-ring resource states~\cite{Bartolucci2021, Bombin2021}, the device must be able to store $6000d^2$ photonic resource states simultaneously. With a maximum delay length of 1 $\mu$s, 4.7 million resource states must be generated every 1 $\mu$s, i.e., a total RSG rate of 4.7 THz is required. Due to a code cycle of 1 $\mu$s, such a device generates keys at the same rate as the superconducting qubit device. With 10 $\mu$s delays, the device footprint can be reduced by a factor of 10 to a 470 GHz total RSG rate, while the time per key increases by a factor of 10 to 1.6 days. Longer delays of 100 ms and 1000 ms decrease the device footprint in additional factor-of-10 steps, and increase the time per key to 16 days and 160 days.

\subsection{Active-volume architecture}

We consider two scenarios for the active-volume architecture: one in which we are generating one key at a time using 3000 logical memory qubits, and one in which we are generating four keys in parallel using 12000 logical memory qubits. In the first scenario, the active volume per key is 6.6 billion blocks, while in the second scenario, the active volume per key is reduced to 3.75 billion blocks due to the cheaper inversion operation.

\textbf{Determining the code distance.} In an active-volume architecture, the total spacetime volume is twice the active volume, i.e., $1.3 \times 10^{10}$ logical blocks in the first scenario. This can also be understood from the observation that 3000 workspace qubits execute 3000 active-volume blocks in every logical cycle, so it will take $2.2 \times 10^6$ logical cycles to finish the computation. The number of logical cycles multiplied by the 6000 logical qubits present in the device also yields a total spacetime volume of $1.3 \times 10^{10}$ logical blocks. If we now choose the code distance such that the total error probability after executing $1.3 \times 10^{11}$ logical blocks is below 50\%, we can reduce the code distance to $d=24$.

\textbf{Circuit-based quantum computers.} While the non-local connections in an active-volume architecture are motivated by photonics, the architecture is hardware-agnostic. We can perform an active-volume estimate for superconducting qubits and trapped ions, even if no concrete proposal for an implementation of these non-local connections exists for these qubits. In the first scenario, the device consists of 6000 qubit modules, each containing a distance-24 surface-code patch, i.e., 1152 physical qubits. Executing 3000 blocks per logical cycle, the device generates one key every 58 seconds (superconducting qubits) or 16 hours (trapped ions), which is 240 times faster with a 27\% smaller footprint compared to the baseline estimate. In the second scenario, the device footprint increases by a factor of 4, but due to the cheaper inversion operation, the time per key decreases by a factor of 7.

\textbf{Photonic FBQC.} We obtain similar resource estimates for photonic fusion-based quantum computers, for which a concrete implementation of an active-volume architecture is described in Ref.~\cite{Litinski2022}. Compared to a baseline architecture, the required RSG rate in the first scenario decreases by 27\% due to the reduced code distance. The time per key decreases by the same factor of 240, where device footprint and computational speed can traded off linearly by varying the delay length. However, the decrease in the cost per key with larger memory capacity implies that we should always seek to maximize the delay length, as the extra logical qubits unlocked by longer delay lines can be used to reduce the computational volume cost per key. In the second scenario, the cost per key is reduced by an additional 40\%. Our results are summarized in Fig.~\ref{fig:summary}.

\subsection{Reaction depth}
\label{sec:reactiondepth}

One important consideration in any resource estimate is the reaction limit. Due to the sequential nature of the classical processing related to surface-code-based FTQC, it is impossible to execute computations faster than the reaction depth multiplied by the reaction time~\cite{Fowler2012a}. The reaction time is the time that it takes to perform a layer of single-qubit measurements, feed the measurement results into a decoder, perform a decoding tasks, and use the result to send back measurement instructions to the quantum computer. Since the $n$-qubit adders considered in this estimate have a reaction depth of $2n$, we can upper bound the reaction depth as twice the Toffoli count. When we are executing $k$ instances of the algorithm in parallel, the subroutines can be parallelized straightforwardly, so the naive estimate of the reaction depth should be divided by $k$.

A typical assumption in the literature is a reaction time of 10 $\mu$s. In this case, the algorithm generating one key at a time would be naively reaction limited at 36 minutes per key, whereas the algorithm generating four keys in parallel would be reaction limited at 5 minutes per key (or 20 minutes for groups of four keys). This is particularly problematic in implementations with short code cycles like superconducting qubits or photonic devices with short delay lines. We have highlighted some estimates in Fig.~\ref{fig:summary} with an asterisk, which are those that would be reaction limited with a 10 $\mu$s reaction time and without further optimizations. Note that, in this case, the reaction limit is irrelevant for photonic devices with long delay lines and slow ion traps.

There are several ways one can avoid the reaction limit. The first is to reduce the reaction time. The assumed 10 $\mu$s are not based on physical limits, and an optimized architecture can, in principle, feature lower reaction times. The second is to use lower-depth subroutines. We have not focused on optimizing the reaction depth. One can use lower depth subroutines for arithmetic operations~\cite{Rines2018,Gidney2020} or depth-reducing tricks such as oblivious carry runways~\cite{Gidney2019a} to increase the cost of the algorithm in favor of a lower depth. The third option is only relevant for photonic FBQC, which is to increase the delay length. Longer delay lines can be used to avoid the reaction limit, as they increase the code cycle time in favor of a smaller footprint. Note that this does not make the device less efficient. On the contrary, the additional logical qubits gained by longer delays make the device more efficient due to the option of cheaper subroutines. If two devices have the same footprint in terms of total RSG rate but different delay lengths, the device with the longer delay length will both generate keys at a faster rate and feature a lower (i.e., less strict) reaction limit. 

\section{Conclusion}

We have performed an active-volume estimate for the cost of generating elliptic curve private keys on a fault-tolerant quantum computer. We also introduced three algorithmic modifications that reduce the Toffoli count per key by up to a factor of 5. We found that this algorithm  benefits particularly strongly from the non-local connections in an active-volume architecture. When generating one key at a time, the active volume architecture features an over 300 times lower cost per key. This can be understood from the observation that the spacetime cost per Toffoli in a baseline architecture is proportional to $4n_Q$ due to the presence of four sequential $T$ gates per Toffoli, whereas in an active-volume architecture it is an $n_Q$-independent cost of $\approx \! 60$ in a computation that primarily relies on arithmetic operations. For $n_Q = 3000$, this results in a factor of 200. The reduced code distance decreases the cost by an additional factor of $(28/24)^3 \approx 1.6$. With more memory qubits to take advantage of the cheaper modular inversion, the cost difference between baseline and active volume architectures can be up to a factor of 700 per key.

Photonic FBQC particularly benefits from long delay lines in this example. Not only do long delay lines increase the memory of device, but they also increase the value that can be extracted from each component, as the additional memory can be used to decrease the cost of some subroutines. Long delay lines also help avoid the reaction limit without sacrificing device performance.

We focused on resource estimates for 256-bit keys at a physical error rate corresponding to 10\% of the surface code threshold error rate. While larger key sizes may benefit from a more careful balancing of window sizes, the rough behavior is that doubling the key size results in a twofold increase in footprint and an eightfold increase in Toffoli count and active volume (and time per key) due to the cubic scaling in key size. A larger physical error rate will also increase the footprint and runtime. Closer to 50\% threshold, we can expect the code distance to roughly double, resulting in a fourfold increase in footprint and a twofold increase in the computational time per key.

Compared to the active-volume estimate for 2048-bit RSA keys in Ref.~\cite{Litinski2022}, the active volume per 256-bit ECC key is lower by a factor of 100-300. However, the active volume per 2048-bit RSA key may be reduced compared to Ref.~\cite{Litinski2022} through further optimizations, e.g., by adjusting the window size of windowed arithmetic to rebalance the contributions of arithmetic and data lookups to the active volume. Still, one can expect the generation of 2048-bit RSA keys to be over an order of magnitude more expensive than 256-bit ECC keys.

\textbf{Open problems.}
We leave several regimes for optimization unexplored in this paper. The first is the low-footprint regime, as we have overestimated the memory requirement with 3000 logical qubits per instance without carefully counting qubits. An active-volume architecture allows for the arbitrary allocation of workspace and memory qubits, so one can slow down the computation in favor of a footprint reduction by an additional factor of up to 2. Furthermore we have not explored the depth optimizations outlined in Sec.~\ref{sec:reactiondepth} that may be required to avoid the reaction limit in implementations with short code cycles. Also, while we have focused on a comparison between general-purpose architectures with strict 2D locality and logarithmic non-local connections, there exist 2D-local special-purpose architectures~\cite{Gidney2019,Gidney2021} that may still perform worse than active-volume architectures, but outperform baseline architectures without requiring non-local connections. One may then investigate how the cost per key can be reduced in a 2D-local implementation. 

More generally, we hope that this work encourages the exploration of similar trade-offs where additional qubits can be used to reduce the cost of subroutines. One similar example described in the literature is Hamming weight phasing~\cite{Gidney2018,Kivlichan2020}, where the cost of commuting equiangular rotations can be reduced substantially by introducing additional ancilla qubits. An application of this technique analogous to the parallel modular inversion would be to execute multiple instances of an algorithm containing arbitrary-angle rotations and then use Hamming weight phasing on groups of rotations from different instances. This reduces the effective cost of rotations in each instance, even if the rotations in any one instance of the algorithm are not part of groups of commuting or equiangular rotations.

In summary, there is motivation to explore additional examples of subroutines that demonstrate reduced costs per instance when executed in parallel, akin to an ``economies of scale'' effect in quantum algorithm design. We hope that this paper will serve as a comprehensive reference on how to perform resource estimation and algorithm optimization for active-volume architectures.

\section*{Acknowledgments}

I would like to thank Naomi Nickerson and Sam Pallister for discussions on the structure of this paper, Sam Roberts for discussions on the below-threshold distance scaling of the logical error rate of surface-code patches, and Dan Dries and Terry Rudolph for helpful feedback on an early draft.

\appendix

\section{Additional figures}

\begin{figure*}[t]
\centering
\includegraphics[width=0.9\linewidth]{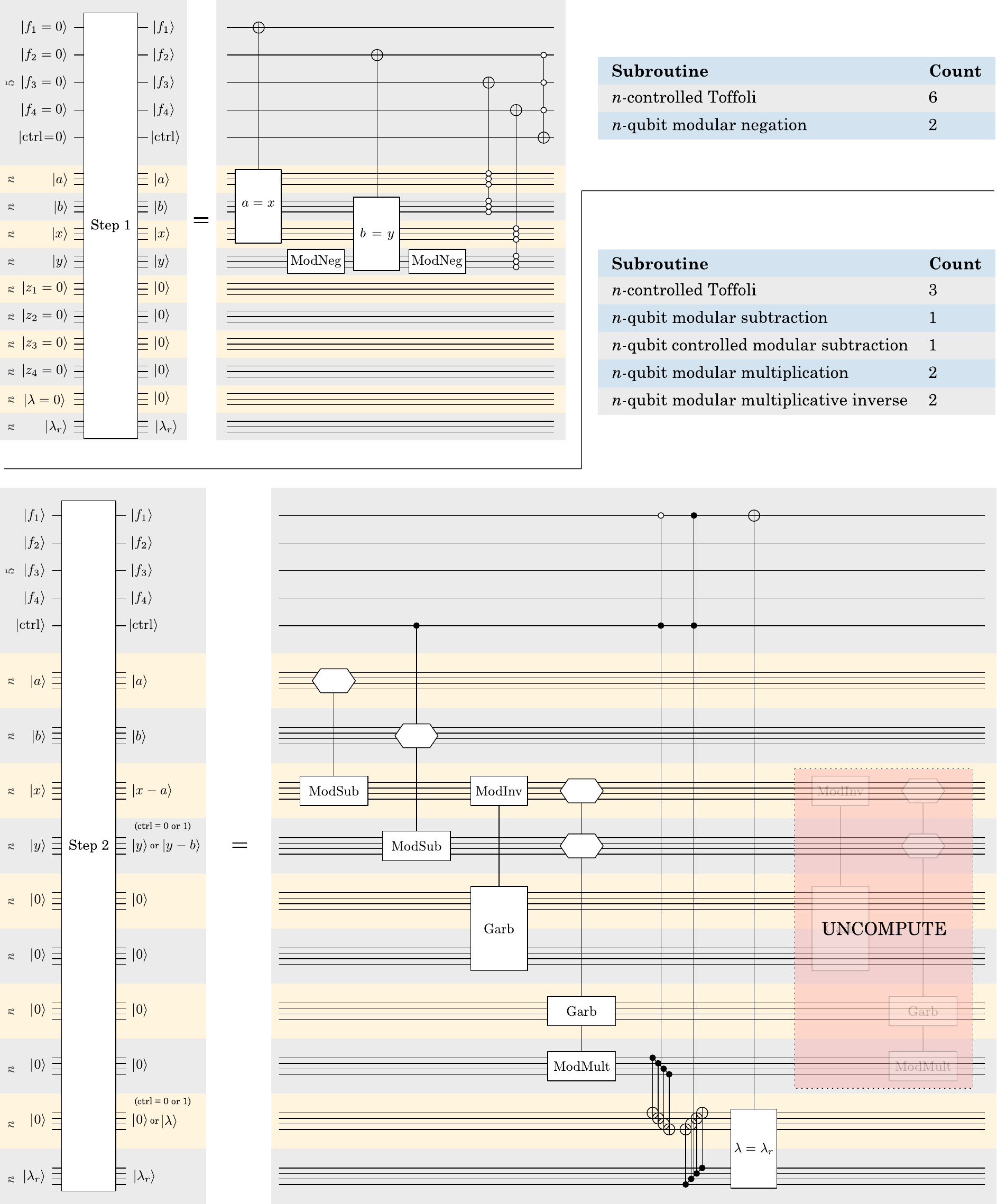}
\caption{Steps 1 and 2 of the elliptic curve point addition circuit in Fig.~\ref{fig:ecpointaddfull}.}
\label{fig:ecpointadd1}
\end{figure*}

\begin{figure*}[t]
\centering
\includegraphics[width=0.96\linewidth]{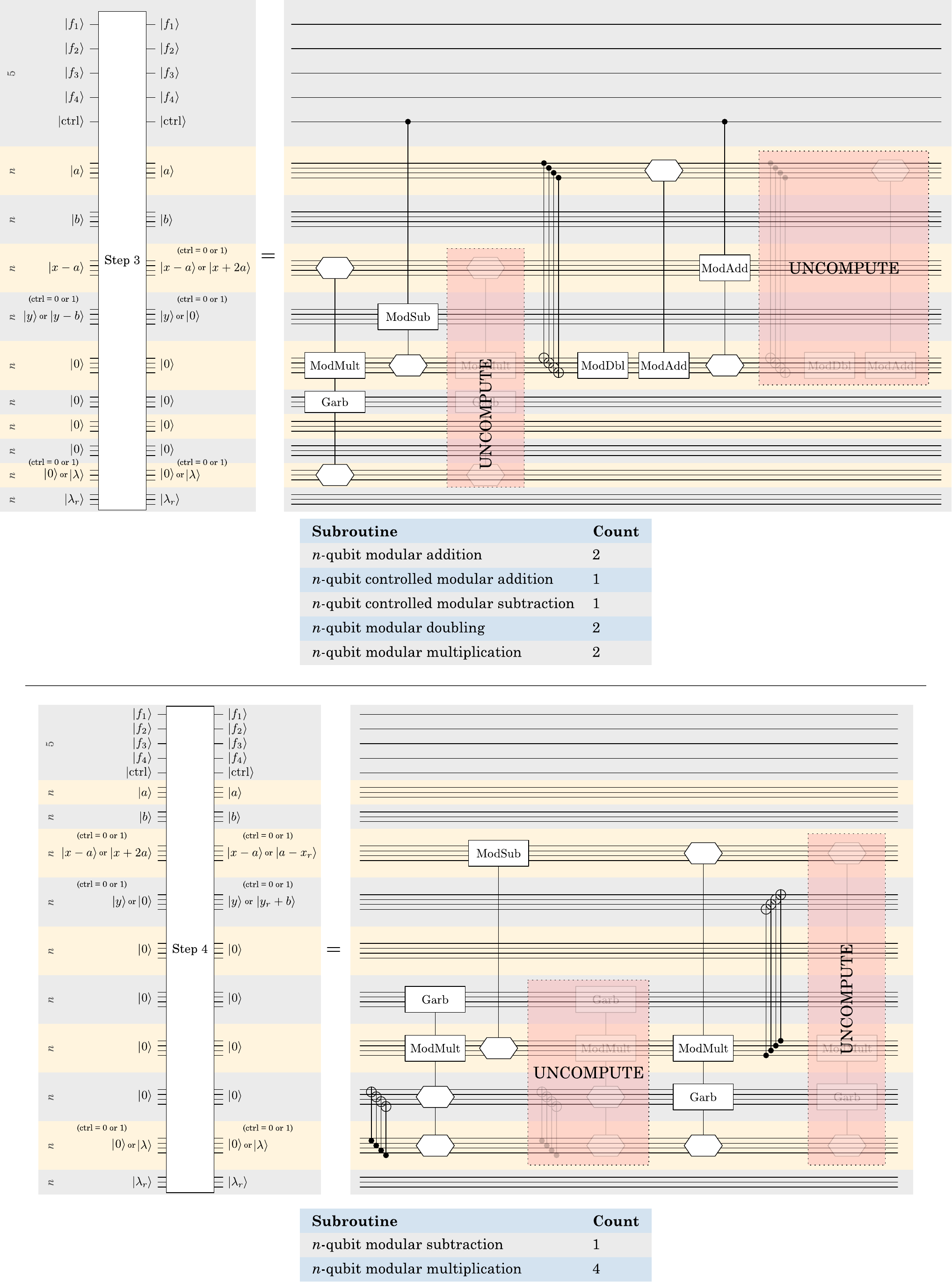}
\caption{Steps 3 and 4 of the elliptic curve point addition circuit in Fig.~\ref{fig:ecpointaddfull}.}
\label{fig:ecpointadd2}
\end{figure*}

\begin{figure*}[t]
\centering
\includegraphics[width=\linewidth]{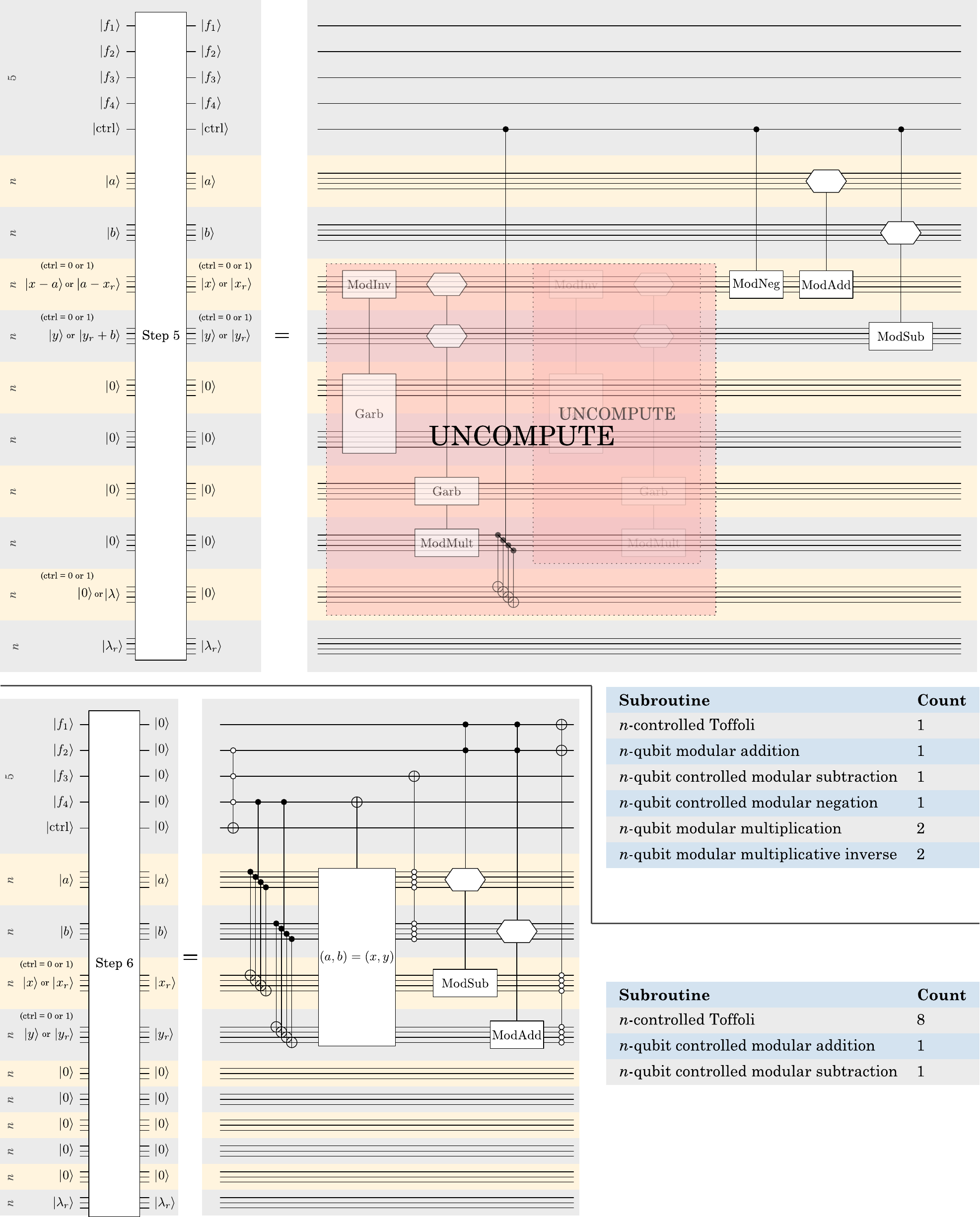}
\caption{Steps 5 and 6 of the elliptic curve point addition circuit in Fig.~\ref{fig:ecpointaddfull}.}
\label{fig:ecpointadd3}
\end{figure*}

\begin{figure*}[t]
\centering
\includegraphics[width=\linewidth]{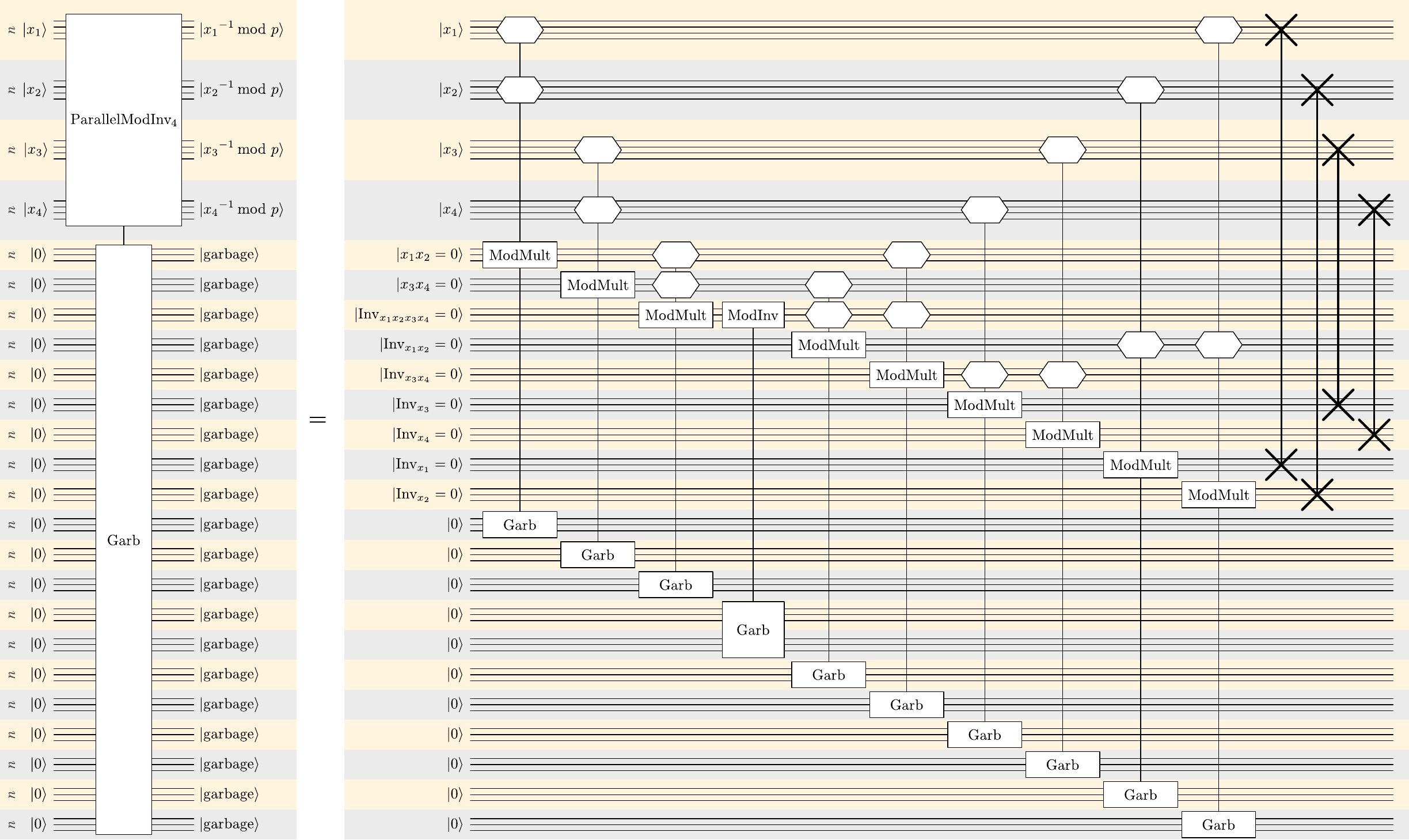}
\caption{Parallel modular inversion of four numbers.}
\label{fig:parallelmodinv1}
\end{figure*}

\begin{figure*}[t]
\centering
\includegraphics[width=\linewidth]{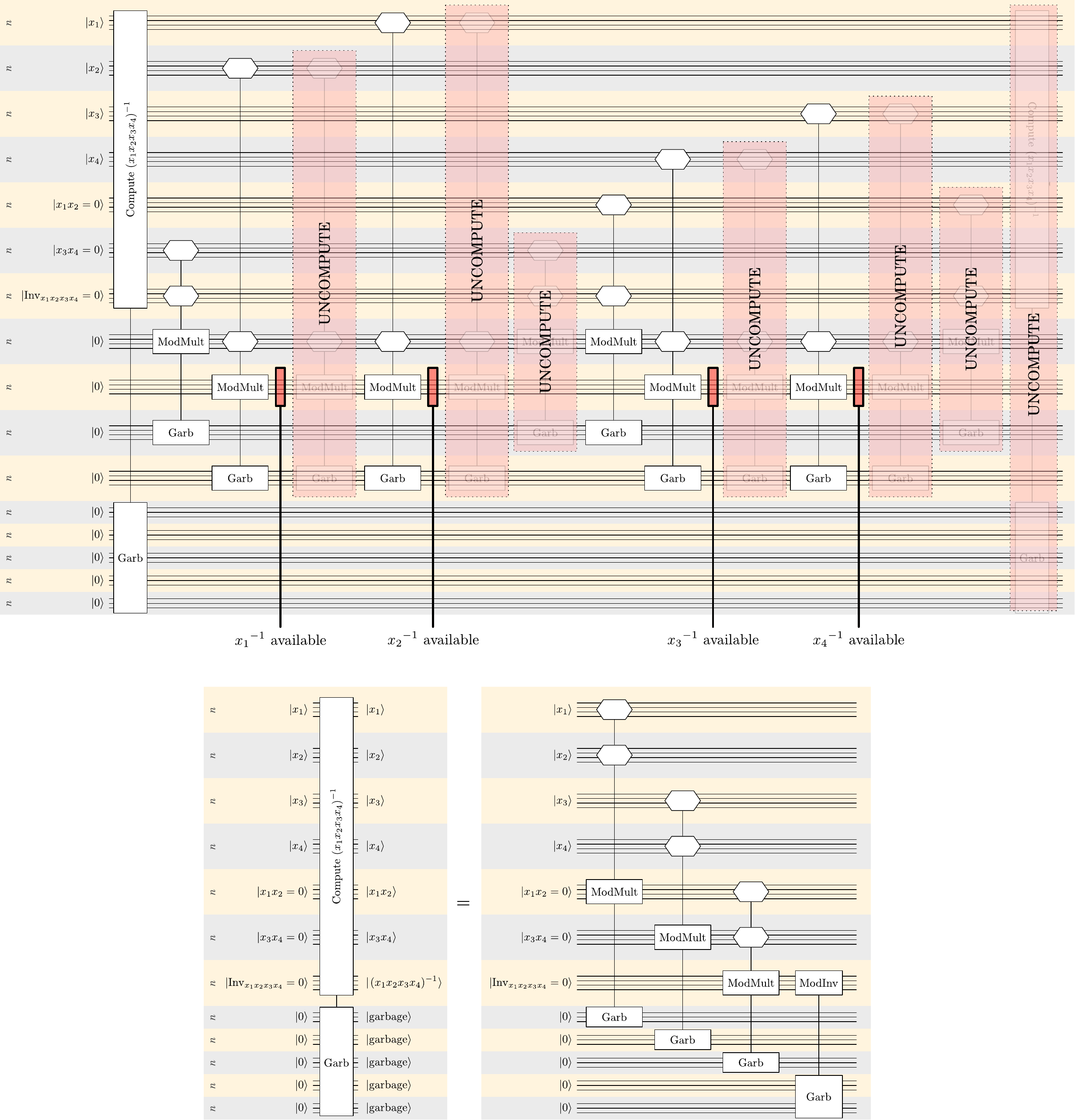}
\caption{Parallel modular inversion of four numbers in which the four inverses are made available sequentially to save qubits.}
\label{fig:parallelmodinv2}
\end{figure*}

Additional figures are shown in Figs.~\ref{fig:ecpointadd1}-\ref{fig:parallelmodinv2}.

\end{document}